\documentclass[prl,twocolumn,english,superscriptaddress,floatfix,longbibliography]{revtex4}

\usepackage{graphicx}
\usepackage{dcolumn}
\usepackage{bm}
\usepackage{xcolor}
\usepackage{amsmath}
\usepackage{amssymb}
\usepackage{stmaryrd}

\newcommand{\ket}[1]{|#1\rangle}
\newcommand{\bra}[1]{\langle#1|}
\newcommand{\ccw}{\circlearrowleft}
\newcommand{\cw}{\circlearrowright}
\newcommand{\ketbra}[2]{|#1\rangle\langle #2|}

\newcommand{\rev}[1]{\textcolor{black}{#1}}
\newcommand{\kk}[1]{\textcolor{black}{#1}}

\newcommand{\kkr}[1]{\textcolor{black}{#1}}

\begin{document}

\title{Topological quantum state control through 
exceptional-point proximity}


\author{Maryam Abbasi}
\affiliation{Department of Physics, Washington University, St.~Louis, Missouri 63130}
\author{Weijian Chen}
\affiliation{Department of Physics, Washington University, St.~Louis, Missouri 63130}
\affiliation{Center for Quantum Sensors, Washington University, St.~Louis, Missouri 63130}
\author{Mahdi Naghiloo}
\affiliation{Department of Physics, Washington University, St.~Louis, Missouri 63130}
\affiliation{Research Laboratory of Electronics, MIT,  Cambridge, Massachusetts 02139}
\author{Yogesh N. Joglekar}
\email[]{yojoglek@iupui.edu}
\affiliation{Department of Physics, Indiana University Purdue University
Indianapolis (IUPUI), Indianapolis, Indiana 46202}
\author{Kater W. Murch}
\email[]{murch@physics.wustl.edu}
\affiliation{Department of Physics, Washington University, St.~Louis, Missouri 63130}
\affiliation{Center for Quantum Sensors, Washington University, St.~Louis, Missouri 63130}
\date{\today}

 \altaffiliation[Also at ]{Physics Department, Washington University in St. Louis.}
\date{\today}

\begin{abstract}

We study the quantum evolution of a non-Hermitian qubit realized as a submanifold of a dissipative superconducting transmon circuit. Real-time tuning of the system parameters to encircle an exceptional point 
results in non-reciprocal quantum state transfer. 
We further observe chiral geometric phases accumulated under state transport, verifying the quantum coherent nature of the evolution in the complex energy landscape and distinguishing between coherent and incoherent effects associated with exceptional point encircling. Our work demonstrates an entirely new method for control over quantum state vectors, highlighting new facets of quantum bath engineering enabled through dynamical 
non-Hermitian control.

\end{abstract}
\maketitle

Small quantum systems that interact with an environment can be described by a Lindblad density matrix equation that encodes their approach to steady state. When the quantum trajectories of these decoherence-inducing dynamics are restricted to those with no quantum jumps, the resulting evolution is described by an effective non-Hermitian Hamiltonian. Such non-Hermitian quantum systems have complex energies, non-orthogonal eigenstates, and undergo a coherent, non-unitary evolution. The presence of special kinds of degeneracies known as exceptional points (EPs) play an important role in the unique characteristics of these non-Hermitian systems \cite{zdem19,Miri19}. Such EPs occur when both the eigenvalues and eigenstates of the system coalesce. A plethora of phenomena associated with EPs have been revealed in classical platforms such as mechanical and optical systems  \cite{guo09,rute10,peng14pt,hoda14, zeun15,Li2019,xiao17}.
In the vicinity of the EP, the shape of the Riemann manifold that describes the complex energies of a non-Hermitian system can lead to fundamentally new phenomena that are not present in their Hermitian counterparts with strictly real energies.  For a second-order EP degeneracy, quasi-static tuning of the Hamiltonian parameters is expected to map one eigenstate, $\ket{\psi_-}$, to the other $e^{i\chi_+} \ket{\psi_+}$, modulo a global phase $\chi_+$. Furthermore, the geometric part of the global phase is expected to be chiral \cite{Demb01, Demb04,Gao2015,Heis99,Mail05,mehri2008}.   Such mode-switch behavior has been demonstrated in classical systems \cite{xu16, Dopp16,Yoon18,zhan18}, yet the extension of such topological control to quantum systems---with no classical counterpart---has remained an outstanding goal in the field \cite{liu2020}. 



Here, we utilize the quantum energy levels of a superconducting circuit  described by an effective non-Hermitian Hamiltonian to study quantum state control in the vicinity of the system's EPs. While our previous work \cite{nagh19} characterized the static properties  of this non-Hermitian system, we now employ dynamical control of the Hamiltonian parameters and observe chiral quantum state transfer when encircling EPs. We further use an auxiliary level of our quantum circuit to verify the coherent nature of this evolution and examine the geometric phases accumulated from quantum state transport. \kkr{These reveal that a $\pi$ phase difference associated with the chirality of the transport persists under non-Hermitian dynamical quantum evolution.} Finally, we exploit state transfer in the limit of fast, closed-loop parameter variation, which goes beyond the slow driving limit demonstrated in previous works
\cite{xu16, Dopp16,Yoon18,zhan18} and reveals a broad parameter range proximity to EPs for successful state transfer.

\begin{figure*}
    \centering
    \includegraphics[width = \textwidth]{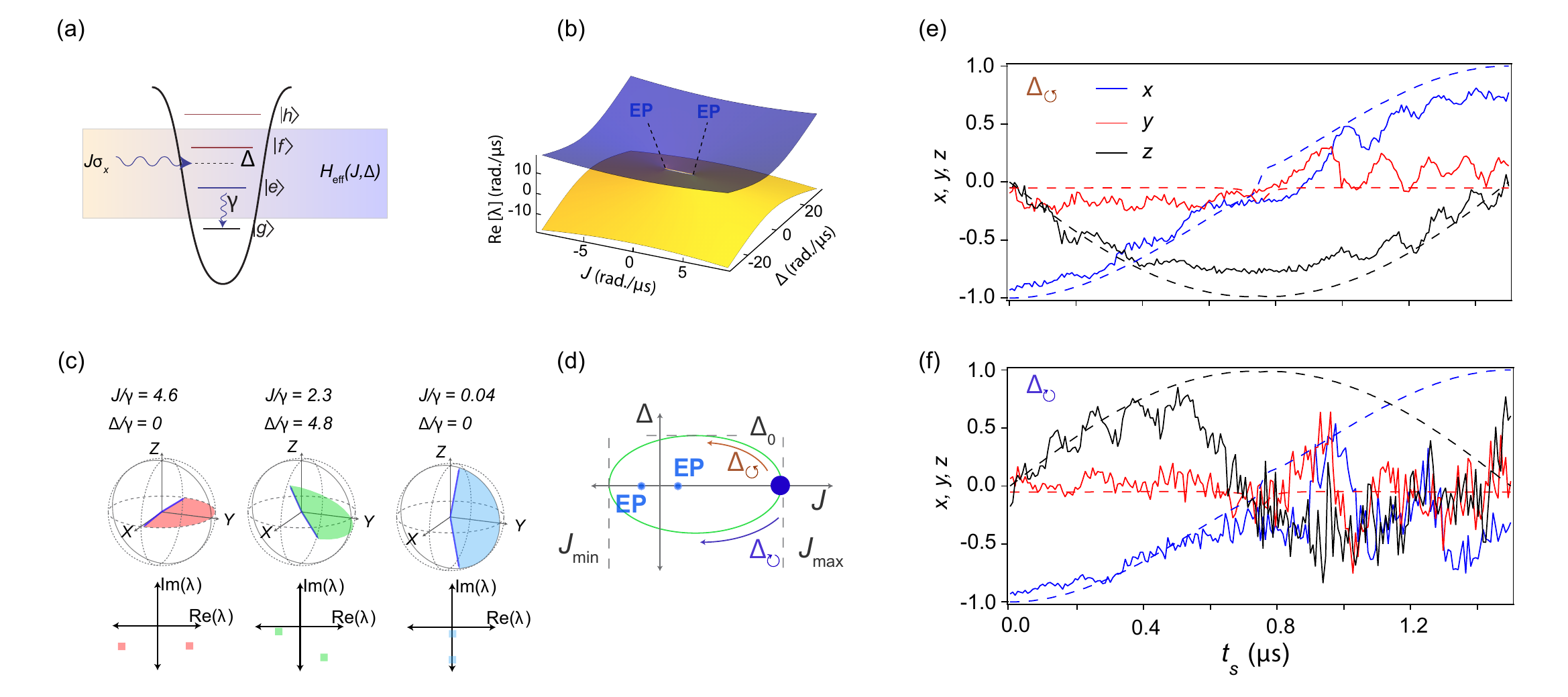}
    \caption{{\bf Dynamically encircling an EP.} (a), The energy states of the transmon circuit with the non-Hermitian qubit submanifold \{$|e\rangle$,$|f\rangle$\} highlighted. The Hamiltonian parameters $J$ and $\Delta$ are tuned with a microwave drive.  (b), In the static limit, the eigenenergies are described by Riemann manifolds. (c), The eigenstates and eigenvalues of $H_\mathrm{eff}$ are indicated for different values of $J$ and $\Delta$. \kkr{The colored planes indicate the opening angle between the two eigenstates for clarity.} {(d)}, The parameter sweeps are designated by direction $(\Delta_\cw,\ \Delta_\ccw)$ and $J_\mathrm{min}$. (e,f), Quantum state tomography \kkr{(solid lines, expressed as the Paul expectation values $x,y,z$}) reveals the state evolution along the parameter path for the $\Delta_\ccw$ ($\Delta_\cw$) direction. The dashed lines indicate the instantaneous eigenstates of $H_\mathrm{eff}$.
  } \label{Fig1}
\end{figure*}

Our experiment comprises a superconducting Transmon circuit \cite{koch07,paik113D} embedded inside a three-dimensional copper cavity (Fig.~\ref{Fig1}a) \cite{supp}. The circuit has anharmonic energy states and the first four energy levels are labeled by $|g\rangle, |e\rangle,$ $|f\rangle$, and $|h\rangle$. The cavity mediates interaction with an environment that is set by the density of states in a microwave transmission line. We shape this density of states to enhance the dissipation of the $|e\rangle$ state while suppressing dissipation of the $|f\rangle$ state. While the evolution of the four-level quantum system can be described by a Lindblad master equation, the evolution within the excited (and lossy) manifold of states $\{|e\rangle, |f\rangle\}$ can be described by an effective non-Hermitian Hamiltonian \cite{Molmer1993,nagh19}. 

By introducing a microwave drive with detuning $\Delta = \omega_{ef}-\omega_{d}$, where $\omega_{ef}$ is the transition frequency between the $|e\rangle$ and $|f\rangle$ states, and $\omega_{d}$ is the microwave drive frequency, we produce the effective Hamiltonian in the frame rotating with the drive:
\begin{equation}
    H_\mathrm{eff}/\hbar= J(|e\rangle \langle f|+|f\rangle \langle e|)+(\Delta-i \gamma /2)|e\rangle \langle e|
\label{Hamiltonian t-independent}
\end{equation}
where $J$ is the coupling rate between $|e\rangle$ and $|f\rangle$, and $\gamma$ is the decay rate of the $|e\rangle$ state. Quantum dynamics of the qubit are given by the (complex) eigenvalues $\lambda_\pm$ and (non-orthogonal) eigenstates $|\psi_\pm\rangle$ of the Hamiltonian \kkr{(expressed in the energy basis)};
\begin{equation}
\lambda_{\pm}=\Delta/2-i \text{$\gamma/4 $}\pm\sqrt{J^2+(\text{$\Delta/2 $}-\text{$i \gamma/4 $})^2},
\label{eigenvalues}
\end{equation}
\begin{equation}
|\psi_\pm\rangle \propto \left(\begin{array}{c} \lambda_\pm\\J \end{array} \right).\label{eignestates2}
\end{equation}
The real part of eigenenergies in the parameter space $(J, \Delta)$ is provided in Fig.~\ref{Fig1}b, and the eigenstates for select values of $J$ and $\Delta$ are sketched in Fig.~\ref{Fig1}c. The static EP degeneracies occur at $J_\mathrm{EP}= \pm \gamma/4 $. 

\begin{figure*}
    \centering
    \includegraphics[width = \textwidth]{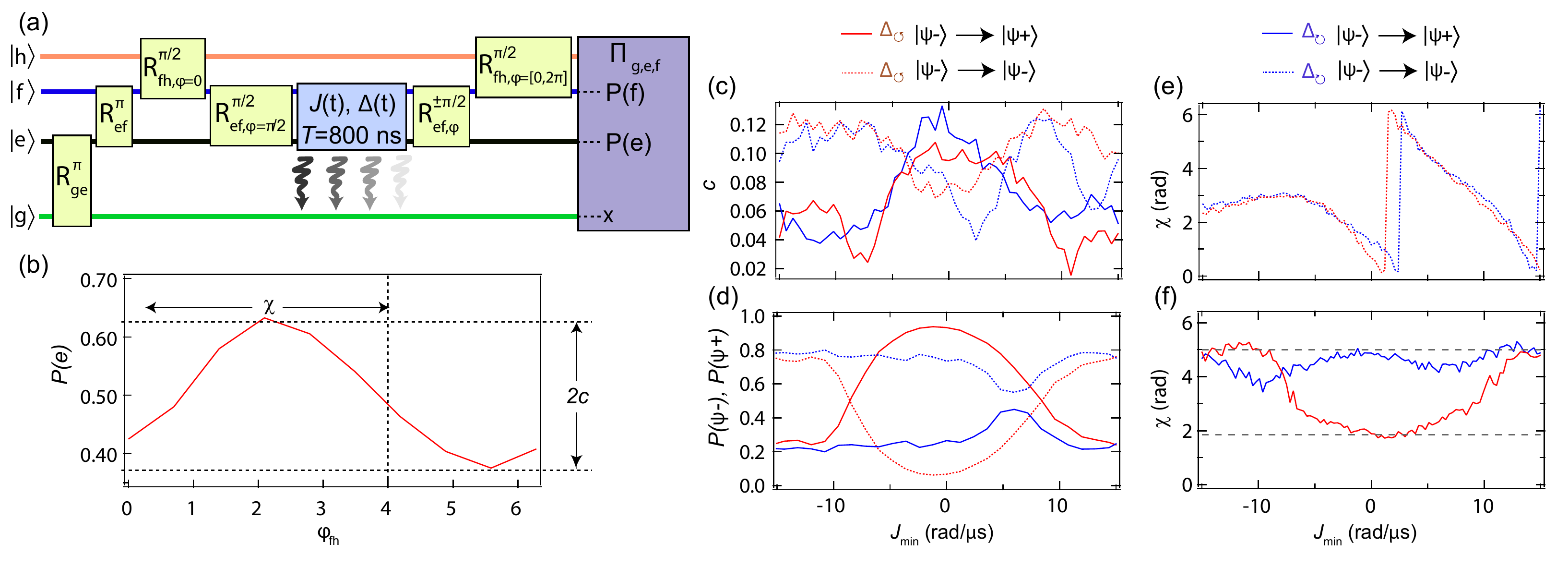}
    \caption{{\bf Coherent state transport and geometric phase measurement}. (a), \rev{Experiment schematic; a series of resonant rotations prepare a superpostion between the state $\ket{\psi_-}$ and state $|h\rangle$.  The $|h\rangle$ state is used as a quantum phase reference to determine the accumulated phase on the quantum states that evolve in the non-Hermitian Floquet Hamiltonian. After evolution for $T=800$ ns, the rotation $R_{ef}^{\pm \pi/2}$ determines which qubit state ($\ket{\psi_+}$ or $\ket{\psi_-}$) is interfered with the $\ket{h}$ reference.} {(b)}, By sweeping the phase of the final $R_{fh}^{\pi/2}$ rotation we determine the contrast $c$ and phase $\chi$. 
    \kk{ The interference contrast (c) and state populations (d) of final states $\ket{\psi_\pm}$ for $\Delta_\cw$ and  $\Delta_\ccw$ sweep directions. (e, f), Extracted total phases; dashed gray lines indicate a phase difference of $\pi$. } }
    \label{phase}
\end{figure*}


\kkr{Quantum state tomography \cite{stef06} allows us to study the state of the qubit as the parameters of the Hamiltonian are tuned in real time. We study a parameter loop specified} by initial/final parameters $\Delta=0$ and $J_\mathrm{max}=30$ rad/$\mu$s and the parameter variation; $\Delta(t)=\Delta_{\ccw,\cw} \sin(2\pi t/T)$ and $J(t) = (J_\mathrm{max}-J_\mathrm{min})\cos^2(\pi t/T) + J_\mathrm{min}$ (Fig.~\ref{Fig1}d). \kkr{The sign of $\Delta_{\ccw,\cw}$ determines whether the sweep is clockwise or counterclockwise, and the sign of $J$ is determined by the phase of the drive. Tomography along the path is achieved by dividing the time evolution into sequentially longer steps $t_\mathrm{s}\in [0,T]$,} pausing the evolution at $t_\mathrm{s}$ and performing measurements to determine Pauli expectation values $x\equiv \langle \sigma_x\rangle$, $y \equiv \langle \sigma_y \rangle $ and $z \equiv \langle \sigma_z \rangle$ \kkr{in the energy basis of the $\{\ket{e},\ket{f}\}$ qubit}.

We slowly vary the system parameters in a loop given by $T=1.5 \ \mu$s and $J_\mathrm{min}= 0.3 \ \mathrm{rad}/\mu$s. By choosing $\Delta_\ccw=10\pi \ \mathrm{rad}/\mu$s (Fig.~\ref{Fig1}e), the system evolves from $\rho_-$, where $\mathrm{tr}(\rho_- \sigma_x)\simeq -1$, roughly following the instantaneous eigenstates of $H_\mathrm{eff}$. After a complete loop that encircles the EP, the system does not return to the initial state, instead the final state is close to $\rho_+$ which is nearly orthogonal to the initial state. This observation can be qualitatively understood by walking through the Riemann structure associated with the static EP; \kkr{the qubit follows the Riemann surface crossing onto the lower sheet at the branch cut connecting the two EPs (Fig.~\ref{Fig1}b).}  In addition, finite time evolution induces transitions between the two eigenstates of the system, leading to a small oscillation in the Pauli expectation values, with frequency given by the real part of the energy difference of the eigenstates.  

In contrast, if we choose $\Delta_\cw=-10 \pi \ \mathrm{rad}/\mu$s, corresponding to encircling the static EP in a clockwise direction, the system does not evolve along the instantaneous eigenstates. As shown in Fig.~\ref{Fig1}f, the state significantly deviates from the eigenstate in the vicinity of the EP.  This can be attributed to non-Hermitian gain/loss effects observed in previous works \cite{xu16, Dopp16,Yoon18,zhan18} as well as other sources of dissipation \cite{fabr2019,chen20}. \kkr{Along this parameter path, the imaginary component of the eigenenergy corresponds to larger loss, resulting in a reduced postselection probability as can be seen in the increased noise in the data. The loss of one eigenstate can be viewed as a relative gain of the other eigenstate. Any small fraction of population that is seeded by non-adiabaticity or dissipation into the relative gain eigenstate is therefore amplified. This gain/loss effect does not occur for the $\Delta_\ccw$ sweep because the system follows the instantaneous eigenstate with relative gain which is stable against non-adiabaticity and dissipation. }



\rev{In order to investigate the quantum nature of \emph{state} transport, as opposed to the transfer of \emph{population} between eigenstates  
\cite{xu16, Dopp16,Yoon18,zhan18}}, we make use of the $\ket{h}$ level as a quantum phase reference, as shown in Fig.~\ref{phase}a. Resonant rotations are used to initialize the three-state system in the state $\rho \propto (\ket{h}+\ket{\psi_-})(\bra{h}+\bra{\psi_-})$. The qubit then undergoes dynamical evolution under the time-dependent Hamiltonian specified by  $J_\mathrm{min}$, $\Delta_{\ccw,\cw} = \pm 10 \pi\ \mathrm{rad}/\mu$s, and for $T = 800$ ns. After this evolution, the three-state system is in general in a mixed state, $\rho \propto c_-(\ket{h}+e^{i\chi_-}\ket{\psi_-})(\bra{h} + e^{-i\chi_-}\bra{\psi_-}) + c_+(\ket{h}+e^{i\chi_+}\ket{\psi_+})(\bra{h} + e^{-i\chi_+}\bra{\psi_+})$ involving both qubit eigenstates, where $\chi_\pm$ are phases accumulated on the states. \rev{We note that coherent terms such as $\ket{\psi_-}\bra{\psi_+}$ remain negligible during the evolution.} A second rotation is used to rotate either the $\ket{\psi_+}$ or $\ket{\psi_-}$ into the state $\ket{f}$ which then interferes with the $\ket{h}$ reference. We determine the contrast $c$ and total phase $\chi$ from the resulting interference (Fig.~\ref{phase}b). 
    
In Figure \ref{phase}c, we display the measured contrast for both final states. In the vicinity of $J_\mathrm{min}=0$, we observe higher contrast for the $\ket{\psi_+}$ final state for both the $\Delta_\ccw$ and $\Delta_\cw$ parameter sweeps, indicating that the state transport ($\ket{\psi_\mp}\to e^{i\chi_\pm} \ket{\psi_\pm}$) is quantum coherent. In comparison, we display in Fig.~\ref{phase}d the relative population in the two states. 
Near $J_\mathrm{min}=0$, our observation of larger populations in the $\ket{\psi_+}$($\ket{\psi_-}$) states for $\Delta_\ccw$($\Delta_\cw$) sweeps is consistent with ``chiral'' features associated nonreciprocal population/energy transfer observed in previous work \cite{xu16, Dopp16,Yoon18,zhan18}. Here relative gain/loss of the two paths favors one or the other final states. This chiral effect, however, is comparatively incoherent, showing reduced contrast despite larger population. 
Whereas the gain/loss effects arising from the imaginary energy components can favor population transfer between states, this process is not necessarily coherent. The interference contrast therefore distinguishes between coherent state transport and incoherent population transfer between states. 

\kkr{We now examine the total quantum phases accumulated for the two encircling directions, as displayed in Fig.~\ref{phase}e,f.} In general, the total quantum phase will be the sum of a dynamical phase arising from Hamiltonian evolution and a geometric phase. This is apparent in Fig.~\ref{phase}e where we observe significant dependence of the phase on the sweep parameter $J_\mathrm{min}$. However, for state transport that follows the Riemann surfaces, we expect the dynamical phase to cancel as the state spends equal time in either energy eigenstate. This results in the relative insensitivity of the total phase to the sweep parameter as shown in Fig.~\ref{phase}f. Here we observe a $\pi$ phase difference between the $\Delta_\ccw$ and $\Delta_\cw$ sweeps as is anticipated from the static structure of EPs \cite{Demb01,Demb04, Gao2015,Heis99,Mail05,mehri2008}. \kkr{Qualitatively, this $\pi$ phase difference arises because one path passes through the excited state of the qubit, while the other does not.}

\begin{figure}[!ht]
    \centering
     \includegraphics[width = 0.8\linewidth]{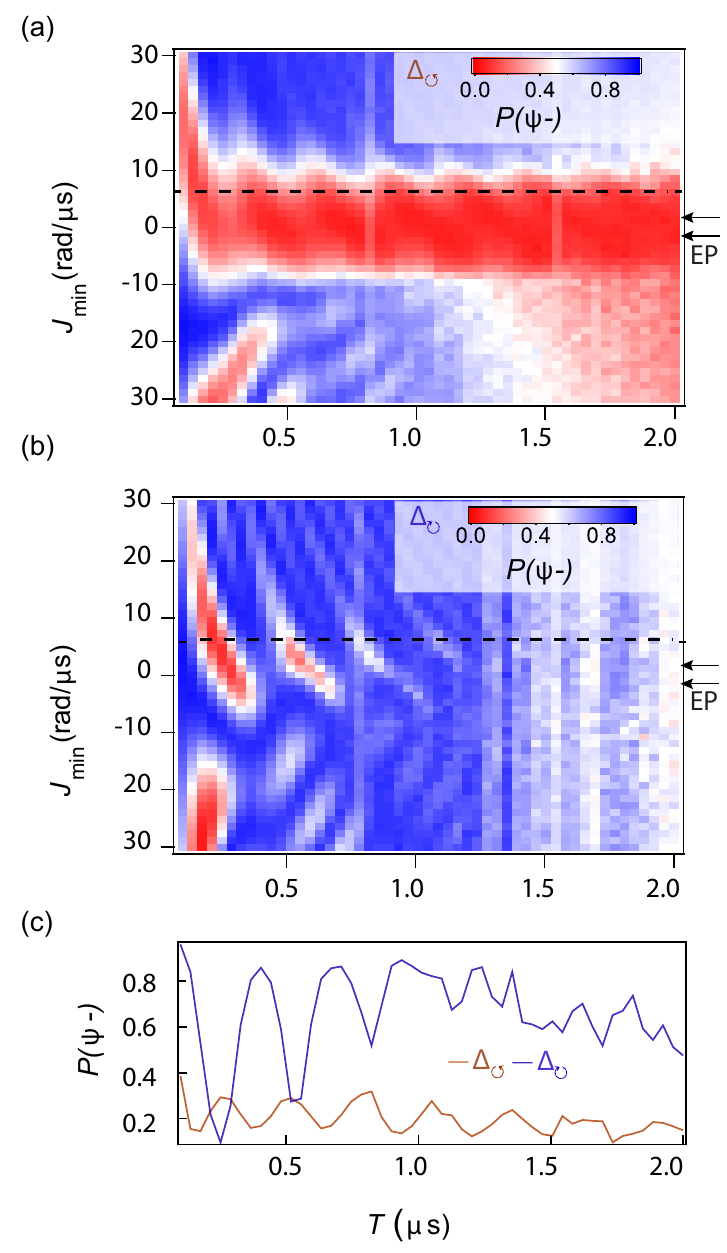}
    \caption{{\bf Population transport beyond slow-driving limit.}  {(a,b)}, The eigenstate population $P(\psi_-)$ after one period evolution for the two sweep directions $\Delta_\ccw$ and $\Delta_\cw$ are displayed versus $J_\mathrm{min}$ and $T$. Black arrows indicating the location of second-order static EPs.
    (c), Two cuts from panels {(a)} [(b)] shown in blue [red] at $J_\mathrm{min}=6 \ \mathrm{rad}/\mu\mathrm{s}$. 
    }
    \label{floquet}
\end{figure}

While the quantum state transfer under quasistatic tuning of the system parameters is best 
understood as a walk through the complex-energy landscape of a static Hamiltonian $H_\mathrm{eff}$ with EPs (Fig.~\ref{Fig1}b), our results in Fig.~\ref{phase}d clearly indicate that in the dynamical case the state transfer can happen in a broad range of the parameter $J_\mathrm{min}$ that includes the situations of encircling zero, one, and two EPs. \kkr{We attribute this observation to the non-adiabatic coupling near the EPs, which occurs when the parameter sweeps is not infinitely slow \cite{Hass2017}. }

\kkr{To further investigate the nature of state transfer beyond the slow driving limit, as is relevant in any real-time operation on quantum states, we now study the population transfer in the limit of fast parameter variation. As before, we prepare the system in the state \rev{$\rho_- \simeq  \ketbra{\psi_-}{\psi_-} $} and perform closed loop parameter variation for different loop periods $T$ and $J_\mathrm{min}$. After one complete encircling (at time $t=T$) we use quantum state tomography to determine $P(\psi_-)$, the population in $\ket{\psi_-}$ as displayed in Fig.~\ref{floquet}.  We consider both $\Delta_\ccw$ (Fig.~\ref{floquet}a) and $\Delta_\cw$ (Fig.~\ref{floquet}b) directions. We observe a rich dependence on the parameters of the loop (Fig.~\ref{floquet}c). 
These observations indicate that successful quantum state transfer can occur in the  fast driving limit.} 

\rev{Our investigation of state transport in the vicinity of exceptional point degeneracies reveals new methods of quantum coherent state control \kkr{of dissipative systems} enabled through non-Hermitian Hamiltonian dynamics. This work, and the robustness with which we observe the predicted chiral geometric phases, opens new avenues to investigations of eigenvalue braiding in larger dimension non-Hermitian systems \cite{carl2019,hoel20}, allowing the study of exotic topological classes of these (knotted) systems.  Future extensions to non-Hermiticities through non-reciprocity \cite{wang19} would enable scaling to quantum many-body systems where the study of topological edge-states and invariants~\cite{yao18,kuns18} are expected to yield deviations from the paradigmatic bulk-boundary correspondence~\cite{Weidemann2020,Xiao2020}. Finally, the interplay of quantum measurement dynamics \cite{Weber2014,Minev2019,rege21} with the non-Hermitian dynamics explored here is expected to produce new fruitful avenues for quantum control.
}
    


\section{Acknowledgements}

We thank J. Harris and C. Bender for discussions. This research was supported by NSF Grant No. PHY- 1752844 (CAREER), AFOSR MURI Grant No. FA9550-21-1-0202,  and the Institute of Materials Science and Engineering at Washington University. 






\pagebreak
	
	\newpage  
	
	\textcolor{white}{.}
	\newpage
	
\widetext

\begin{center}
	\textbf{\large Supplemental Material for ``Topological quantum state control through exceptional-point proximity''}
\end{center}

In the Supplementary Materials, we provide detailed description of our experimental setup and protocol, theoretical modeling, and comparison of experimental results and theoretical calculations.

\section{I. Experimental Setup and protocol}

\emph{Experimental setup}---\rev{The Transmon transition frequencies are
$\omega_{ge}/2\pi= 5.7$ GHz, 
$\omega_{ef}/2\pi= 5.41$ GHz, 
$\omega_{fh}/2\pi= 5.08$ GHz.}  We isolate the dynamics described by the non-Hermitian Hamiltonian by using high fidelity single shot readout to eliminate evolution that carries the system out of the excited manifold of states.  The readout is achieved through standard dispersive measurement techniques \cite{wall05}, with coupling rates for the $|e\rangle$ and $|f\rangle$ states given by $\chi_e/2\pi=-2$ MHz, and $\chi_f/2\pi=-11$ MHz, respectively. \kk{The dispersive interaction results in a qutrit-state-dependent phase shift on a resonant cavity probe at frequency $\omega_c/2\pi = 6.684$ GHz which is subsequently amplified with a Josephson parametric amplifier \cite{hatr11para,cast08} operating in phase sensitive mode. Accessing the populations of all three levels enables us to post-select on the experimental sequences where the evolution remains in the $\{|e\rangle, |f\rangle\}$ manifold. } \rev{The final measurement fidelities are approximately $98\%$ for $|f\rangle$, $80\%$ for $|e\rangle$, and $82\%$ for $|g\rangle$. For the tomography data displayed in Fig.~\ref{Fig1}e,f we exclusively use measurement on the $|f\rangle$ state to improve the fidelity of the postselection. }

\emph{Quantum state tomography}---\kk{ Quantum state tomography is achieved by pausing the evolution and performing measurements along the $X$, $Y$ and $Z$ axes of the qubit. Sampling $~10^{4}$ measurements along each axis yields the qubit Pauli expectation values $x\equiv \langle \sigma_x\rangle$, $y \equiv \langle \sigma_y \rangle $ and $z \equiv \langle \sigma_z \rangle$. Measurements about the different axes are obtained by abruptly setting $J,\Delta=0$ and performing rotations about the $X$ and $Y$ axes, or no rotation, followed by a measurement along the $Z$ axis. For all measurements, we discard any experimental sequence where the final measurement finds the qubit has left the $\{|e\rangle,|f\rangle\}$ manifold. }

\emph{Geometric phase}---\rev{When the parameters of a Hamiltonian are tuned in a closed-loop fashion, the eigenstates acquire a combination of geometric and dynamical phases. In our experiment we prepare a superposition of the state $\ket{\psi_-}$ and the quantum phase reference state $\ket{h}$ using microwave pulses:  we apply a resonant $\pi$ rotation to transfer all population from $|g\rangle$ to $|e\rangle$; followed immediately by a second $\pi$ rotation from $|e\rangle$ to $|f\rangle$; a $\pi/2$ rotation on the $\{|h\rangle,|f\rangle\}$ manifold  prepares the state $\propto (|h\rangle+|f\rangle)$; finally, a $\pi/2$ rotation along the $Y$ axis of the qubit ($\{|e\rangle,|f\rangle\}$)  manifold results in an equal superposition of $|h\rangle$ and $|\psi_-\rangle$ \kkr{(since $\Delta=0$ and $J\gg J_\mathrm{EP}$)}. Following this state preparation, we apply a drive to the qubit to produce the Hamiltonian $H_\mathrm{eff}$. The parameters ($\Delta(t)$, $J(t)$) of $H_\mathrm{eff}$ are tuned over a period of $T=800$ ns.}  

\rev{Based on Fig.~\ref{phase}d we anticipate that the final populations in $\ket{\psi_\pm}$ will depend on the sweep parameter $J_\mathrm{min}$. Two primary effects contribute to this dependence: first, the Riemann surface for the real part of the eigenenergies of the static Hamiltonian predicts a state mapping behavior $\ket{\psi_-}\to \ket{\psi_+}$ for $|J_\mathrm{min}|<J_\mathrm{EP}$, independent of sweep direction, and second, the imaginary energy component yields a loop direction dependent effect, as displayed in Fig.~\ref{Fig1}e,f. }

\rev{After the parameter sweep is concluded we utilize a resonant qubit $\pi/2$ rotation to selectively rotate the $\ket{\psi_-}$ or $\ket{\psi_+}$ state into $\ket{f}$. A final $\pi/2$ rotation on the $\{|h\rangle,|f\rangle\}$ manifold completes the Ramsey measurement.  By stepping the phase of this final rotation, we shift the interference by one complete ``fringe'', allowing the measurement of the phase offset $\chi$ and contrast. The final projective measurement $\Pi_{g,e,f}$ is composed of a $\pi$ rotation on the qubit manifold to transfer population in $\ket{f}$ to the state $\ket{e}$ before readout. }

\rev{This interference measurement allows us to determine the interference phase for four possible interference paths (two directions, two final states). The two paths with $\ket{\psi_-}\to \ket{\psi_-}$, for $\Delta_\cw$ and  $\Delta_\ccw$ accumulate total phases that are a combination of dynamical and geometric phase. In contrast, for the two paths $\ket{\psi_-}\to \ket{\psi_+}$, for $\Delta_\cw$ and  $\Delta_\ccw$, the state spends equal time in either eigenstate, causing the dynamical phase to cancel.}

\section{II. Theoretical Modeling for Coherent State Transport and Geometric Phase measurement}

To theoretically model the dynamics of the system, we utilize the Lindblad master equation:
\begin{equation}
   \frac{ \partial \rho(t)}{\partial t} = - \frac{i}{\hbar} [H_{c}(t),\rho(t)]+ \sum_{i=1}^{4} L_i \rho(t) L_i^{\dagger}- \frac{1}{2}\{L_i^{\dagger}L_i,\rho(t)\}
 \end{equation}
 where $H_c/\hbar= J(|e\rangle\langle f|+|f\rangle\langle e|) +\Delta/2(|e\rangle\langle e|-|f\rangle\langle f|)$ and $L$ is the jump operator for relaxation and dephasing for 4 quantum state system. The relaxation operator includes $\sqrt{\gamma_e }|g\rangle\langle e|$, $\sqrt{\gamma_f}  |e\rangle\langle f|$, and $\sqrt{\gamma_h } |h\rangle\langle f|$. The dephasing contribution to Lindblad operator includes $\sqrt{\gamma_{2e }/2} |e\rangle\langle e|$, $\sqrt{\gamma_{2f }/2} |f\rangle\langle f|$, and $\sqrt{\gamma_{2h }/2} |h\rangle\langle h|$. The dissipation rates are: $\gamma_e=\rev{6.2}/\mu$s, $\gamma_f=0.32/\mu$s, 
$\gamma_h=0.36 /\mu$s,  
$\gamma_{2e}=3.7/\mu$s,
$\gamma_{2f}=\rev{0.9}/\mu$s, 
$\gamma_{2h}=1.4/\mu$s.

\begin{figure}[!h]
     \centering
    \includegraphics[width = .6 \textwidth]{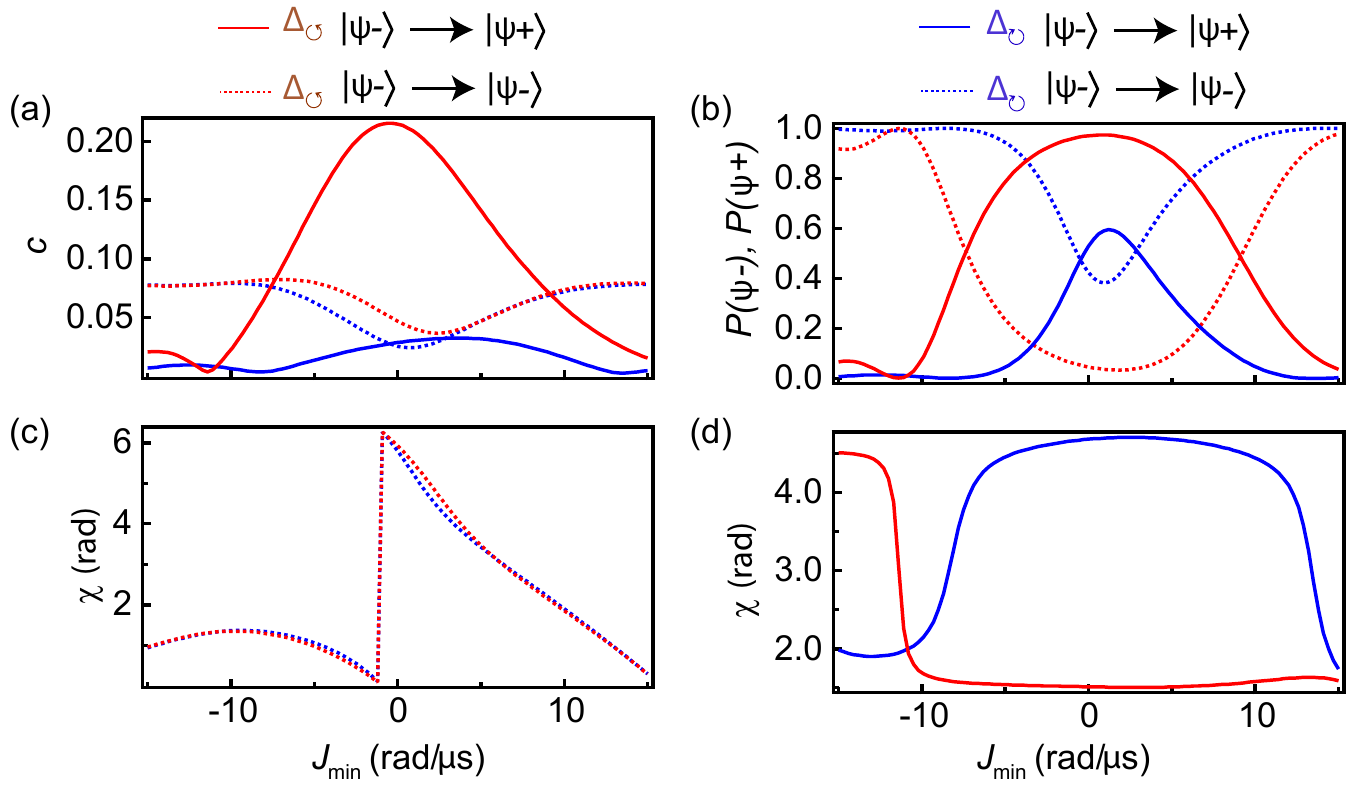}
    \caption{Theoretical calculations of the interference contrast (a), state population (b), and the accumulated phase (c,d)  of the final states $|\psi_\pm\rangle$ for both encircling directions. The dissipation only comes from the relaxation of the $|e\rangle$ level. Parameters used for the theoretical calculations are: $\gamma_e=\rev{6.2}/\mu$s, $J_\mathrm{max}=30$ rad/$\mu$s, $\Delta=10\pi$ rad/$\mu$s. }
    \label{phase_simulation_g_e}
\end{figure}

In order to form an understanding of the role of different dissipation channels in our observations of interference contrast presented in the main text, we now model the system  in the presence of different dissipation sources. First, we only consider relaxation of the $|e\rangle$ level with the jump operator $\sqrt{\gamma_e} |g\rangle\langle e|$. In this case, the evolution in the $\{|e\rangle, |f\rangle \}$ submanifold obtained through postselection is described by an effective non-Hermitian Hamiltonian (Eq.~\ref{Hamiltonian t-independent}). We examine the population transport and quantum state coherence during encircling EPs at different $J_\mathrm{min}$ and observe that the population of the final state qualitatively matches with the coherent contrast [Fig.~\ref{phase_simulation_g_e}(a,b)]. Further, we clearly observe a $\pi$ phase difference between the final states $|\psi_+\rangle$ at two encircling directions [Fig.~\ref{phase_simulation_g_e}(d)] that is insensitive to the values of $J_\mathrm{min}$.


\begin{figure}[!h]
     \centering
    \includegraphics[width = .6 \textwidth]{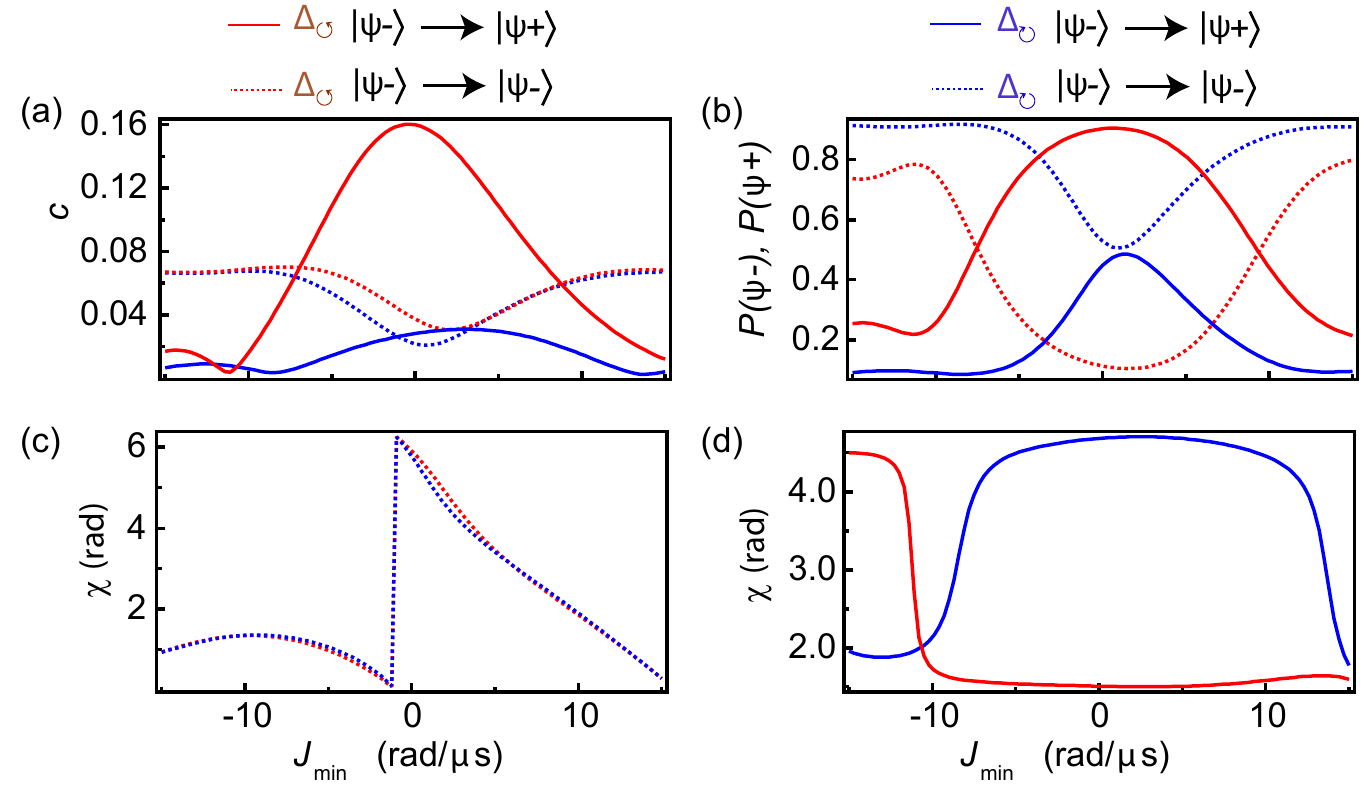}
    \caption{Theoretical calculations of the interference contrast (a), state populations (b), and the accumulated phase (c,d)  of the final states $|\psi_\pm\rangle$ for both encircling directions. The dissipation comes from relaxation of both the $|e\rangle$ and $|f\rangle$ levels as well as pure dephasing of the $\{|e\rangle, |f\rangle\}$ submanifold. Parameters used for the theoretical calculations are: $\gamma_e=\rev{6.2}/\mu$s, $\gamma_f=0.32/\mu$s, 
    $\gamma_{2f}=\rev{0.9}/\mu$s, $J_\mathrm{max}=30$ rad/$\mu$s, $\Delta=10\pi$ rad/$\mu$s.
    }
    \label{phase_simulation_g_ef2}
\end{figure}


Next, we include the dissipation of spontaneous emission and pure dephasing of the $\{|e\rangle, |f\rangle \}$ submanifold in the theoretical modeling [Fig.~\ref{phase_simulation_g_ef2}]. We note that the accumulated phases during the dynamics are not affected by these additional dissipation sources as shown in Fig.~\ref{phase_simulation_g_ef2}(c,d). However, the population transport shows a different trend from the coherent contrast for the clockwise encircling direction. Though the population transport suggests the eigenstate $|\psi_-\rangle$ dominates at the end, this process is not necessarily coherent and shows less coherent contrast compared to the other eigenstate $|\psi_+\rangle$ (compare the relative heights of the blue solid and dashed curves in Fig.~\ref{phase_simulation_g_ef2}(a,b)). These observations qualitatively agree with our experimental observations shown in Fig.~\ref{phase} of the main text.

\begin{figure}[!h]
    \includegraphics[width = 1 \textwidth]{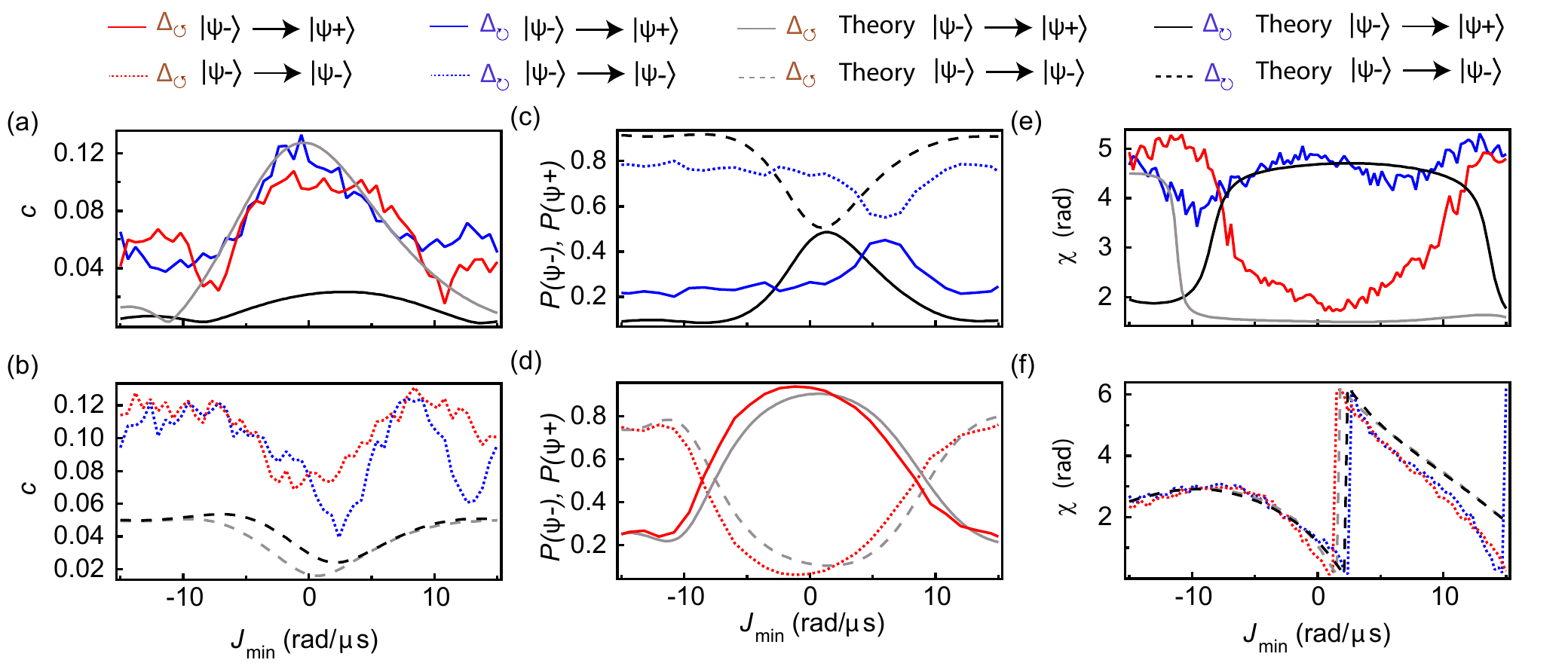}
    \caption{Comparison between theoretical calculations and experimental results of the interference contrast (a,b), state population (c,d), and the accumulated phase (e,f)  of the final states $|\psi_\pm\rangle$ for both encircling directions. The blue and red curves are the experimental results, and the black and gray curves are the theoretical results. Parameters used for the theoretical calculations are: $\gamma_e=\rev{6.2}/\mu$s, $\gamma_f=0.32/\mu$s,
    $\gamma_h=0.36 /\mu$s,  
    $\gamma_{2f}=\rev{0.9}/\mu$s, $J_\mathrm{max}=30$ rad/$\mu$s, $\Delta=10\pi$ rad/$\mu$s. 
    }
    \label{phase_simulation}
\end{figure}

To best describe our experiments, we then include the dissipations of the $|h\rangle$ level into our modeling. Figure~\ref{phase_simulation} compares the theoretical results with the experimental observations shown in Fig.~\ref{phase} of the main text.  We note reasonable qualitative agreement between the Lindblad calculations and the experimental data. 
Residual qualitative and quantitative disagreement between the calculations and measurements arise from effects not included in the Lindblad modeling such as the readout fidelity and the effects of charge noise associated with higher energy levels.

\begin{figure*}
    \centering
    \includegraphics[width = \textwidth]{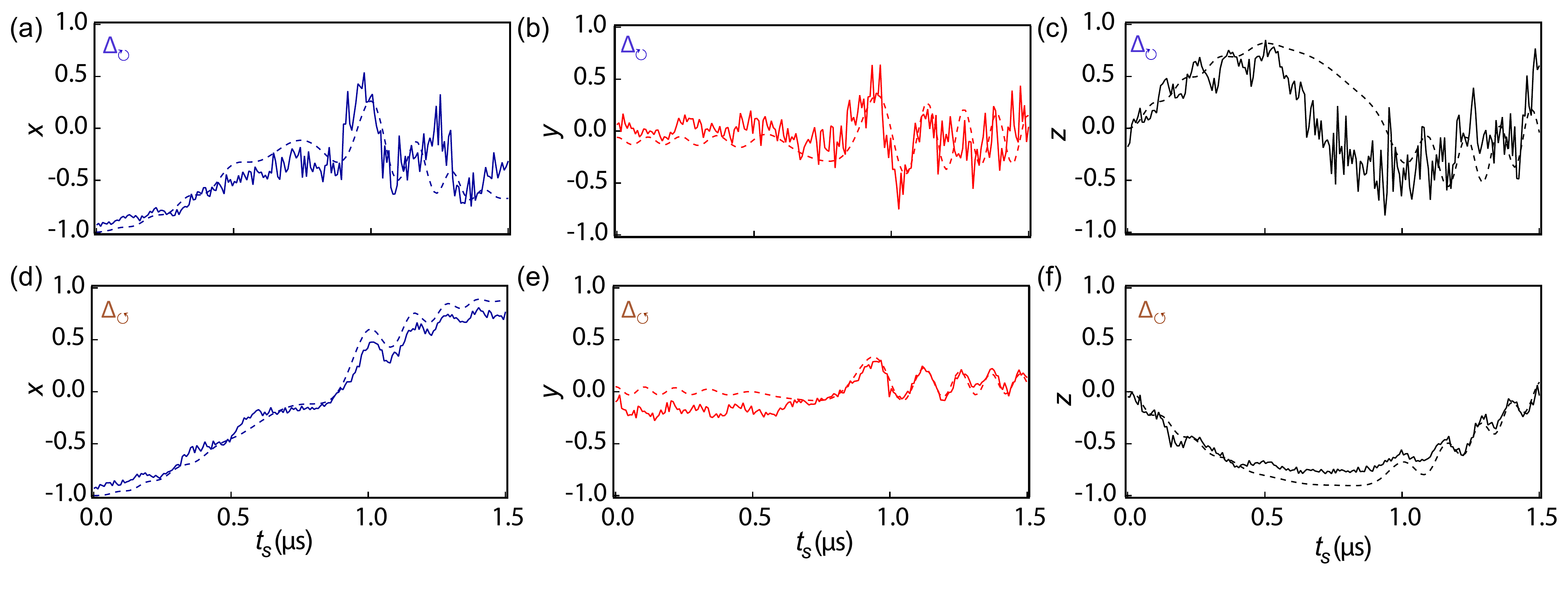}
    \caption{Comparision of experimental results of quantum state tomography (solid curves) and theoretical predictions from Lindblad master equation (dashed curves) when dynamically encircling an EP. 
    The parameters used in the calculations are: $\gamma_e=\rev{6.2}/\mu$s, $\gamma_f=0.32/\mu$s, $J=30$rad/$\mu$s, and $\Delta=10$ rad/$\mu$s.}
    \label{simulation}
\end{figure*}

In Fig.~\ref{simulation} we provide a comparison to the calculated Bloch vector components using the Lindblad master equation to the data provided in Fig.~\ref{Fig1}(e,f).  We note close agreement between the experimental tomography and the calculations. 

\begin{figure}
    \centering
     \includegraphics[width = 0.6\textwidth]{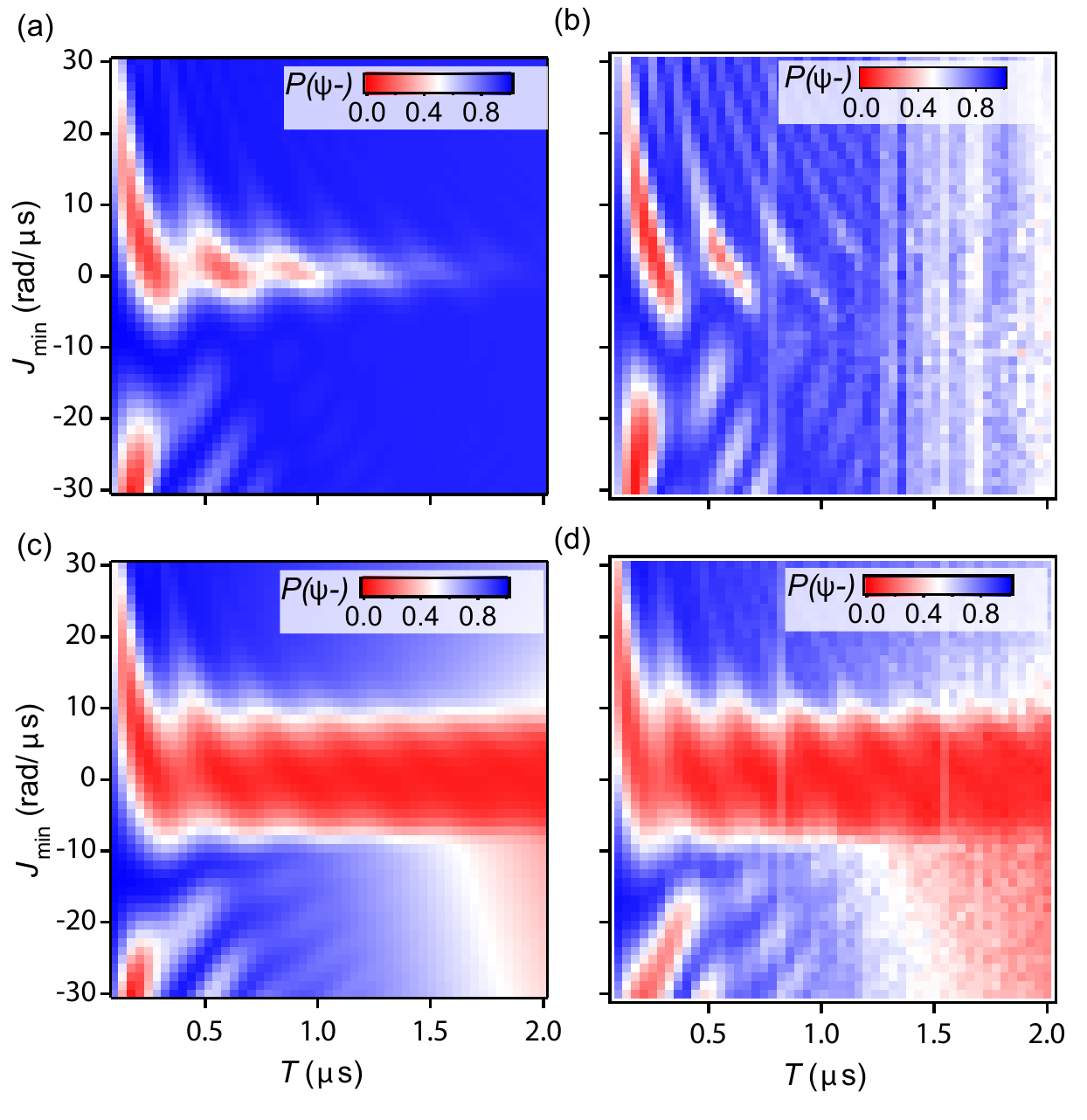}
    \caption{
    Theoretical calculations (left) and experimental results (right) of the eigenstate population $P(\psi_-)$ after one period evolution for the two sweep directions $\Delta_\ccw$ (a,b) and $\Delta_\cw$ (c,d) are displayed versus $J_\mathrm{min}$ and $T$.
  }
    \label{fig:EPExpVSSimFloquet}
\end{figure}

Finally, Fig.~\ref{fig:EPExpVSSimFloquet} displays comparisons between the population transport from the Lindblad calculation and the experimental data from Fig.~\ref{floquet}.  We note that many of the rich details of the data are reproduced by the calculation.



\end{document}